\documentclass[12pt]{article}

\usepackage{amsmath,epsf,amssymb,latexsym}
\usepackage[ps,dvips,matrix,arrow,frame,import,curve,color]{xy}
\xyoption{ps}
\xyoption{dvips}

\def\FIGa{{
\begin{figure}[!tbp]
\[
\begin{xy} <0.5mm,0mm>:
(-10,0)*{}="v1", (10,0)*{}="v2", (40,0)*{+}, (60,10)*{}="v3", (60,-10)*{}="v4",
(-11,0)*{}="v1g",(11,0)*{}="v2g",
(-24,14)*{}="e1", (-24,-14)*{}="e2", (24,-14)*{}="e3", (24,14)*{} ="e4",
(46,24)*{}="e5", (46,-24)*{}="e6", (74,-24)*{}="e7", (74,24)*{}="e8",
\ar@{-}|*\dir{>}"v1";"e1",
\ar@{-}|*\dir{>}"e2";"v1",
\ar@{~}"v1g";"v2g",
\ar@{-}|*\dir{>}"v2";"e3",
\ar@{-}|*\dir{<}"v2";"e4",
\ar@{-}|*\dir{<}"e5";"v3",
\ar@{-}|*\dir{>}"e8";"v3",
\ar@{~}"v3";"v4",
\ar@{-}|*\dir{>}"e6";"v4",
\ar@{-}|*\dir{<}"e7";"v4"
\end{xy}
\]
\caption{Tree-level Bhabha scattering diagrams.}
\label{fig:bhabha}
\end{figure}
}}

\def\FIGb{{
\begin{figure}[!tbp]
\[
\begin{xy} <0.6mm,0mm>:
 (-40,0)*\xybox{ <0.3mm,0mm>:
 (-10,10)*{}="v1", (-10,-10)*{}="v2", (-24,24)*{}="e1", (-24,-24)*{}="e2", (4,-24)*{}="e3", (4,24)*{}="e4",
\ar@{-}"e1";"v1",
\ar@{-}"e2";"v2",
\ar@{-}"v2";"v1",
\ar@{-}"e3";"v2",
\ar@{-}"e4";"v1"
 }, 
 (-20,0)*{+},
(0,0)*\xybox{ <0.3mm,0mm>:
 (-10,10)*{}="v1", (-10,-10)*{}="v2", (10,-10)*{}="v3", (10,10)*{}="v4", 
 (-24,24)*{}="e1", (-24,-24)*{}="e2", (24,-24)*{}="e3", (24,24)*{}="e4",
\ar@{-}"e1";"v1",
\ar@{-}"e2";"v2",
\ar@{-}"e3";"v3",
\ar@{-}"e4";"v4",
\ar@{-}"v1";"v2",
\ar@{-}"v2";"v3",
\ar@{-}"v3";"v4",
\ar@{-}"v4";"v1"
},
(20,0)*{+},
(40,0)*\xybox{ <0.3mm,0mm>:
 (-10,10)*{}="v1", (-10,-10)*{}="v2", (20,-10)*{}="v3", (20,10)*{}="v4", (5,-10)*{}="v5", (5,10)*{}="v6", 
 (-24,24)*{}="e1", (-24,-24)*{}="e2", (34,-24)*{}="e3", (34,24)*{}="e4",
\ar@{-}"e1";"v1",
\ar@{-}"e2";"v2",
\ar@{-}"e3";"v3",
\ar@{-}"e4";"v4",
\ar@{-}"v1";"v2",
\ar@{-}"v2";"v3",
\ar@{-}"v3";"v4",
\ar@{-}"v4";"v1",
\ar@{-}"v5";"v6"
},
(80,0)*{+~~~~\ldots},
(-45,-30)*\xybox{ <0.3mm,0mm>:
(-10,0)*{}="v1", (10,0)*{}="v2", (-11,0)*{}="v1g",(11,0)*{}="v2g",
(-24,14)*{}="e1", (-24,-14)*{}="e2", (24,-14)*{}="e3", (24,14)*{} ="e4",
\ar@{-}"v1";"e1",
\ar@{-}"e2";"v1",
\ar@{-}"v1g";"v2g",
\ar@{-}"v2";"e3",
\ar@{-}"v2";"e4"
},
(-20,-30)*{+},
(0,-30)*\xybox{ <0.3mm,0mm>:
 (-10,10)*{}="v1", (-10,-10)*{}="v2", (10,-10)*{}="v3", (10,10)*{}="v4", 
 (-24,24)*{}="e1", (-24,-24)*{}="e2", (24,-24)*{}="e3", (24,24)*{}="e4",
\ar@{-}"e1";"v1",
\ar@{-}"e2";"v2",
\ar@{-}"e3";"v3",
\ar@{-}"e4";"v4",
\ar@{-}"v1";"v2",
\ar@{-}"v2";"v3",
\ar@{-}"v3";"v4",
\ar@{-}"v4";"v1"
},
(20,-30)*{+},
(40,-30)*\xybox{ <0.3mm,0mm>:
 (-10,10)*{}="v1", (-10,-10)*{}="v2", (20,-10)*{}="v3", (20,10)*{}="v4", (-10,0)*{}="v5", (20,0)*{}="v6", 
 (-24,24)*{}="e1", (-24,-24)*{}="e2", (34,-24)*{}="e3", (34,24)*{}="e4",
\ar@{-}"e1";"v1",
\ar@{-}"e2";"v2",
\ar@{-}"e3";"v3",
\ar@{-}"e4";"v4",
\ar@{-}"v1";"v2",
\ar@{-}"v2";"v3",
\ar@{-}"v3";"v4",
\ar@{-}"v4";"v1",
\ar@{-}"v5";"v6"
},
(80,-30)*{+~~~~\ldots}
\end{xy}
\]
\caption{Partial double-counting when adding $s$- and $t$-channel Bethe-Salpeter diagrams.}
\label{fig:bethe}
\end{figure}
}}

\def\FIGc{{
\begin{figure}[!tbp]
\[
\begin{xy} <0.3mm,0mm>:
 (-20,20)*{}="v1", (-20,-20)*{}="v2", (20,-20)*{}="v3", (20,20)*{}="v4",
 (-34,34)*{}="e1", (-34,-34)*{}="e2", (34,-34)*{}="e3", (34,34)*{}="e4",
 (-10,-20)*{}="u1",(0,-20)*{}="u2", (10,-20)*{}="u3",
 (-10,20)*{}="u6",(0,20)*{}="u5", (10,20)*{}="u4",
 (-20,-10)*{}="w1",(-20,0)*{}="w2", (-20,10)*{}="w3",
 (20,-10)*{}="w6",(20,0)*{}="w5", (20,10)*{}="w4",
 (0,-40)*{}="t1", (0,-30)*{}="t2"; (5,-35)*{t},
 (-40,0)*{}="s1", (-30,0)*{}="s2"; (-35,6)*{s},
 \ar@{>}"t1";"t2",
 \ar@{>}"s1";"s2",
 \ar@{-}"e1";"v1",
\ar@{-}"e2";"v2",
\ar@{-}"e3";"v3",
\ar@{-}"e4";"v4",
\ar@{-}"v1";"v2",
\ar@{-}"v2";"v3",
\ar@{-}"v3";"v4",
\ar@{-}"v4";"v1",
\ar@{-}"u1";"u6",
\ar@{-}"u2";"u5",
\ar@{-}"u3";"u4"
\ar@{-}"w1";"w6",
\ar@{-}"w2";"w5",
\ar@{-}"w3";"w4"
\end{xy}
\]
\caption{Full double-counting when adding $s$- and $t$-channel planar fishnet diagrams.}
\label{fig:fishnet}
\end{figure}
}}

\def\FIGcc{{
\begin{figure}[!tbp]
\[
\begin{xy} <0.75mm,0mm>:
(-24,24)*{}="u1", (-10,10)*{} ="u2", (10,10)*{}="u3", (24,24)*{}="u4",
(-24,-24)*{}="v1", (-10,-10)*{} ="v2", (10,-10)*{}="v3", (24,-24)*{}="v4",
(-26,21)*{}="w1", (-12,7)*{} ="w2", (-12,-7)*{}="w3", (-26,-21)*{}="w4",
(26,21)*{}="x1", (12,7)*{} ="x2", (12,-7)*{}="x3", (26,-21)*{}="x4",
 (0,-35)*{}="t1", (0,-25)*{}="t2"; (5,-30)*{t},
 (-35,0)*{}="s1", (-25,0)*{}="s2"; (-30,5)*{s},
\ar@{>}"t1";"t2",
\ar@{>}"s1";"s2",
\ar@{-}|*\dir{>}"u1";"u2",
\ar@{-}|*\dir{>}"u2";"u3",
\ar@{-}|*\dir{>}"u3";"u4",
\ar@{-}|*\dir{<}"v1";"v2",
\ar@{-}|*\dir{<}"v2";"v3",
\ar@{-}|*\dir{<}"v3";"v4",
\ar@{-}|*\dir{<}"w1";"w2",
\ar@{-}|*\dir{<}"w2";"w3",
\ar@{-}|*\dir{<}"w3";"w4",
\ar@{-}|*\dir{>}"x1";"x2",
\ar@{-}|*\dir{>}"x2";"x3",
\ar@{-}|*\dir{>}"x3";"x4"
 
\end{xy}
\]
\caption{A duality diagram.}
\label{fig:duality}
\end{figure}
}}

\def\FIGd{{
\begin{figure}[!tbp]
\[
\begin{xy} <1.0mm,0mm>:
 (-10,-10)*{}="s1", (-5,-10)*{}="s2"; (-7,-7)*{s},
 (30,-25)*{}="t1", (30,-20)*{}="t2"; (33,-23)*{t},
 (0,0)*\xybox{ <1.0mm,0mm>:
 (0,0)*{}="v1", (20,0)*{}="v2", (40,0)*{}="v3", (60,0)*{}="v4",
\ar@{-}|*\dir{>}"v1";"v2",
\ar@{-}|*\dir{>}"v2";"v3",
\ar@{-}|*\dir{>}"v3";"v4"
},
(0,-3)*\xybox{<1.0mm,0mm>:
 (0,0)*{} = "v1", (20,0)*{}="v2", (20,-10)*{}="v3", (0,-10)*{}="v4",
\ar@{-}|*\dir{<}"v1";"v2",
\ar@{-}|*\dir{<}"v2";"v3",
\ar@{-}|*\dir{<}"v3";"v4"
},
(23,-3)*\xybox{<1.0mm,0mm>:
 (0,0)*{} = "v1", (14,0)*{}="v2", (14,-10)*{}="v3", (0,-10)*{}="v4",
\ar@{-}|*\dir{<}"v1";"v2",
\ar@{-}|*\dir{<}"v2";"v3",
\ar@{-}|*\dir{<}"v3";"v4",
\ar@{-}|*\dir{<}"v4";"v1"
},
(40,-3)*\xybox{<1.0mm,0mm>:
 (0,0)*{} = "v1", (20,0)*{}="v2", (20,-10)*{}="v3", (0,-10)*{}="v4",
\ar@{-}|*\dir{<}"v1";"v2",
\ar@{-}|*\dir{>}"v1";"v4",
\ar@{-}|*\dir{>}"v4";"v3"
},
(0,-16)*\xybox{ <1.0mm,0mm>:
 (0,0)*{}="v1", (20,0)*{}="v2", (40,0)*{}="v3", (60,0)*{}="v4",
\ar@{-}|*\dir{>}"v1";"v2",
\ar@{-}|*\dir{>}"v2";"v3",
\ar@{-}|*\dir{>}"v3";"v4"
},
(0,-19)*\xybox{ <1.0mm,0mm>:
 (0,0)*{}="v1", (20,0)*{}="v2", (40,0)*{}="v3", (60,0)*{}="v4",
\ar@{-}|*\dir{>}"v1";"v2",
\ar@{-}|*\dir{>}"v2";"v3",
\ar@{-}|*\dir{>}"v3";"v4"
},
\ar@{>}"t1";"t2",
\ar@{>}"s1";"s2"
 \end{xy}
\]
\caption{Planar loop meson-baryon scattering duality diagram.}
\label{fig:planar}
\end{figure}
}}

\def\FIGe{{
\begin{figure}[!tbp]
\[
\begin{xy} <1.0mm,0mm>:
 (-4,-1)*{\text{\bf 1}}, (-4,-16)*{\text{\bf 2}}, (64,-1)*{\text{\bf 4}}, (64,-16)*{\text{\bf 3}},
 (-10,-10)*{}="s1", (-5,-10)*{}="s2"; (-7,-8)*{s},
 (30,-25)*{}="t1", (30,-20)*{}="t2"; (33,-23)*{t},
 (0,0)*\xybox{ <1.0mm,0mm>:
 (0,0)*{}="v1", (20,0)*{}="v2", (40,0)*{}="v3", (60,0)*{}="v4",
\ar@{-}|*\dir{>}"v1";"v2",
\ar@{-}|*\dir{>}"v2";"v3",
\ar@{-}|*\dir{>}"v3";"v4"
},
(0,-3)*\xybox{<1.0mm,0mm>:
 (0,0)*{} = "v1", (10,0)*{}="v2", (10,-10)*{}="v3", (0,-10)*{}="v4",
 (23,0)*{} = "u1", (37,0)*{}="u2", (37,-10)*{}="u3", (23,-10)*{}="u4",
 (50,0)*{} = "w1", (60,0)*{}="w2", (60,-10)*{}="w3", (50,-10)*{}="w4",
\ar@{-}|*\dir{<}"v1";"v2",
\ar@{-}@/_0.5mm/|*\dir{<}"v2";"u4",
\ar@{-}@/^0.75mm/|*\dir{<}"v3";"u1",
\ar@{-}|*\dir{<}"u4";"u3"
\ar@{-}|*\dir{<}"u1";"u2",
\ar@{-}|*\dir{<}"w1";"w2",
\ar@{-}|*\dir{<}"v4";"v3",
\ar@{-}|*\dir{<}"u1";"u2",
\ar@{-}@/_0.5mm/|*\dir{<}"u2";"w4",
\ar@{-}@/^0.75mm/|*\dir{<}"u3";"w1",
\ar@{-}|*\dir{<}"w4";"w3"
},
(0,-16)*\xybox{ <1.0mm,0mm>:
 (0,0)*{}="v1", (20,0)*{}="v2", (40,0)*{}="v3", (60,0)*{}="v4",
\ar@{-}|*\dir{>}"v1";"v2",
\ar@{-}|*\dir{>}"v2";"v3",
\ar@{-}|*\dir{>}"v3";"v4"
},
(0,-19)*\xybox{ <1.0mm,0mm>:
 (0,0)*{}="v1", (20,0)*{}="v2", (40,0)*{}="v3", (60,0)*{}="v4",
\ar@{-}|*\dir{>}"v1";"v2",
\ar@{-}|*\dir{>}"v2";"v3",
\ar@{-}|*\dir{>}"v3";"v4"
},
\ar@{>}"t1";"t2",
\ar@{>}"s1";"s2"
 \end{xy}
\]
\caption{Non-planar loop Pomeron diagram for meson-baryon scattering. }
\label{fig:mesbar}
\end{figure}
}}

\def\FIGf{{
\begin{figure}[!tbp]
\[
\begin{xy} <1.0mm,0mm>:
 (-4,-1)*{\text{\bf 1}}, (-4,-13)*{\text{\bf 2}}, (64,-1)*{\text{\bf 4}}, (64,-13)*{\text{\bf 3}},
 (-10,-7)*{}="s1", (-5,-7)*{}="s2"; (-7,-5)*{s},
 (30,-25)*{}="t1", (30,-20)*{}="t2"; (33,-23)*{t},
 (0,0)*\xybox{ <1.0mm,0mm>:
 (0,0)*{}="v1", (20,0)*{}="v2", (40,0)*{}="v3", (60,0)*{}="v4",
\ar@{-}|*\dir{>}"v1";"v2",
\ar@{-}|*\dir{>}"v2";"v3",
\ar@{-}|*\dir{>}"v3";"v4"
},
(0,-3)*\xybox{<1.0mm,0mm>:
 (0,0)*{} = "v1", (10,0)*{}="v2", (10,-10)*{}="v3", (0,-10)*{}="v4",
 (23,0)*{} = "u1", (37,0)*{}="u2", (37,-10)*{}="u3", (23,-10)*{}="u4",
 (50,0)*{} = "w1", (60,0)*{}="w2", (60,-10)*{}="w3", (50,-10)*{}="w4",
\ar@{-}|*\dir{<}"v1";"v2",
\ar@{-}@/_0.5mm/|*\dir{<}"v2";"u4",
\ar@{-}|*\dir{<}"u4";"u3"
\ar@{-}|*\dir{<}"u1";"u2",
\ar@{-}@/^0.75mm/|*\dir{<}"u3";"w1",
\ar@{-}|*\dir{<}"w1";"w2",
\ar@{-}|*\dir{<}"v4";"v3"
\ar@{-}@/^0.75mm/|*\dir{<}"v3";"u1",
\ar@{-}|*\dir{<}"u1";"u2",
\ar@{-}@/_0.5mm/|*\dir{<}"u2";"w4",
\ar@{-}|*\dir{<}"w4";"w3"
},
(0,-16)*\xybox{ <1.0mm,0mm>:
 (0,0)*{}="v1", (20,0)*{}="v2", (40,0)*{}="v3", (60,0)*{}="v4",
\ar@{-}|*\dir{>}"v1";"v2",
\ar@{-}|*\dir{>}"v2";"v3",
\ar@{-}|*\dir{>}"v3";"v4"
},
\ar@{>}"t1";"t2",
\ar@{>}"s1";"s2"
 \end{xy}
\]
\caption{Non-planar loop Pomeron diagram for meson-meson scattering. }
\label{fig:mesmes}
\end{figure}
}}

\def\FIGg{{
\begin{figure}[!tbp]
\[
\begin{xy} <1.0mm,0mm>:
 (0,0)*\xybox{<1.0mm,0mm>:
 (0,0)*{}="v1", (10,0)*{}="v2", (5,2)*{\text{\bf 1}},
 \ar@{-}"v1";"v2"
 },
 (20,0)*\xybox{<1.0mm,0mm>:
 (0,0)*{}="v1", (10,0)*{}="v2", (5,2)*{\text{\bf 4}},
 \ar@{-}"v1";"v2"
 },
 (10,-5)*\xybox{<1.0mm,0mm>:
 (0,0)*{}="v1", (10,0)*{}="v2",
 \ar@{-}"v1";"v2"
 },
 (15,-10)*\cir<4mm>{},
 (10,-15)*\xybox{<1.0mm,0mm>:
 (0,0)*{}="v1", (10,0)*{}="v2",
 \ar@{-}"v1";"v2"
 },
 (0,-20)*\xybox{<1.0mm,0mm>:
 (0,0)*{}="v1", (10,0)*{}="v2", (5,-2)*{\text{\bf 2}},
 \ar@{-}"v1";"v2"
 },
 (20,-20)*\xybox{<1.0mm,0mm>:
 (0,0)*{}="v1", (10,0)*{}="v2", (5,-2)*{\text{\bf 3}},
 \ar@{-}"v1";"v2"
 },
 (15,-27)*{}="t1", (15,-23)*{}="t2"; (17,-25)*{t},
 \ar@{>}"t1";"t2",
 \end{xy}
\]
\caption{String picture of the diagram of Figure \protect{\ref{fig:mesmes}}. }
\label{fig:string}
\end{figure}
}}
 
\begin{document}
\pagenumbering{arabic}

 \medskip \begin{center} \Large {\bf Two-component Duality and Strings \footnote{Based on
 remarks at the Conference on the Early History of String Theory at the Galileo Galilei Institute of Theoretical Physics in Arcetri, Florence, Italy, on May 19, 2007
 }}\\
  \normalsize                                 
 \bigskip \bigskip                   Peter G.O. Freund\footnote{freund@theory.uchicago.edu}\\
 \medskip \em{Enrico Fermi Institute and Department of Physics\\
 University of Chicago, Chicago, IL 60637}                                 
 							
 \bigskip \bigskip \bigskip
 
 {\bf{Abstract}}\\
 \end{center}

A phenomenologically successful two-component hadronic duality picture led to Veneziano's amplitude, the fundamental first step to string theory. This picture is briefly recalled and its two components are identified as the open strings (mesons and baryons) and closed strings (Pomeron).

\bigskip \bigskip  \bigskip \bigskip

The Veneziano model \cite{V}, the starting point of string theory, addressed the at that point much studied and phenomenologically successful idea of {\em two-component duality}. Here I would like to recall this idea and give its meaning in modern terms. At the risk of letting the cat out of the bag at too early a stage, let me right away say that the two components in question will turn out to be the open and the closed hadronic strings.

The argument for two-component duality is the following. Unlike quarks, hadrons (mesons and baryons) are obviously not elementary objects. Yet the particles appearing in the initial, final and intermediate states of the hadronic $S$-matrix are precisely these composite objects and not their constituent quarks. With elementary particles it is obvious how to calculate the $S$-matrix, just use the Feynman rules. For instance when studying tree-level Bhabha-scattering in QED, we are instructed to {\em add} the $s$- and $t$-channel photon-pole diagrams, as in Fig. \ref{fig:bhabha}. 
\FIGa
But there we have a lagrangian and the photon is an "elementary" particle whose field appears in this lagrangian. When dealing with composite states, these are represented by an infinite sum of Feynman diagrams in the quantum field theory theory of the elementary fields out of which the composite particles are built. In the simplest case we can think of these diagrams as Bethe-Salpeter ladder-diagrams. If only one kind of line (field) )is involved in these ladder-diagrams, as for instance with a $g\Phi^3$ lagrangian, then adding the diagrams in which the composite particle pole appears in the $s$- and in the $t$- channels, would lead to the double-counting of the one-box diagram as in Fig. \ref{fig:bethe}.
\FIGb
If instead of Bethe-Salpeter ladder-diagrams, we would be summing over all planar "fishnet"-diagrams of say a scalar QFT with cubic and quartic self interaction theory (an example of such a planar fishnet diagram is given in Fig. \ref{fig:fishnet}), then not only the one-box diagram, but each and every diagram would be counted twice. This partial, or total double-counting is the essential difference between a theory of elementary and of composite particles.
\FIGc 

In their seminal paper, Dolen, Horn and Schmid \cite{DHS} have shown that for the composite hadrons studied in the laboratory a full such double counting would be involved. I distinctly remember the excitement with which we learned of their paper from Murray Gell-Mann, when he visited us in 1967 before the circulation of the preprint. Dolen, Horn and Schmid carefully showed that in hadronic scattering processes such as $\pi N$- charge-exchange, both the smooth $t$-channel Regge exchange and the bumpy $s$-channel resonances account for the {\em full} amplitude. This is possible because the imaginary part of the $t$-channel (Regge) exchange, averages over the contributions of the absorptive parts of the direct $s$-channel resonances. Therefore the two should {\em not} be added, to avoid double-counting. Rather they are {\em dual} to each other. 

But even for elastic scattering amplitudes for which the $s$-channel is devoid of resonances such as $\pi^+\pi^+, pp, K^+p$, etc... there are Regge-exchanges and for them there is also diffraction with the corresponding Pomeron-exchange. How can this be? The imaginary part of the Regge-exchange comes from the term $e^{-i\pi\alpha(t))}$ in the Regge signature factor. But this term has opposite signs for even- and odd-signature Regge-exchanges. If these are to average to zero, to match the absence of resonances in the $s$-channel, then the even and odd-signature hadronic Regge poles must come in degenerate pairs with matching residue-functions. This {\em exchange degeneracy} is observed experimentally. For instance, extrapolating the rectilinear $\rho$ Regge trajectory to spin 2, it does indeed go through the $f$-meson point, as required by exchange degeneracy. 

But all this still does not take the Pomeron into account. As a consequence of unitarity, diffraction should correspond not to tree-level, but to higher order processes and it was conjeectured by Freund \cite{F} and by Harari \cite{H1} that, unlike the other Regge poles, the Pomeron is dual not to $s$-channel resonances, but to $s$-channel non-resonant background. With this FH-conjecture a two-component picture has thus emerged, in which besides the the mesonic and baryonic Regge trajectories dual to resonances, there is a second component, the Pomeron, dictated by unitarity as a largely $t$-channel flavor singlet trajectory, dual to non-resonant $s$-channel background. This two-component picture accounted for a vast body of data. The remaining question was how to account for the crucial features of the Pomeron this way. 

Mesonic Regge poles and their dualities were modeled by Rosner \cite{R} and Harari \cite{H2} with what were called "duality diagrams" --- such as the one in Fig.\ref{fig:duality} ---  and what are in retrospect clearly open string diagrams, Chan-Paton \cite{CP} rules and all. For the Pomeron loop diagrams are dictated by unitarity. The simplest planar diagram of Fig. \ref{fig:planar} is clearly not of the right type, for neither does it select the flavor singlet in its  $t$-channel, as its dominant part, nor does it correspond to non-resonant background in the $s$-channel.
\FIGcc
\FIGd
 In fact it is nothing more than a loop correction to $t$-channel Regge pole exchange, or equivalently to direct $s$-channel resonances. Freund, Rivers and Jones \cite{FR}, \cite{FJR} found the diagram which selects the flavor singlet to be the one in Fig. \ref{fig:mesbar}. 
\FIGe
It has no three-quark intermediate state in the $s$-channel, in other words no resonances and as such only non-resonant background as required by the FH-conjecture. Moreover, this diagram obviously selects the $t$-channel singlet: the quarks of the hadrons labeled 1 and 4 in Fig. \ref{fig:mesbar} are always the same, as are those for hadrons 2 and 3. 

But the Pomeron is not an exact flavor singlet: the $\pi N$ and $KN$ total cross sections observed in experiment are not equal. The beauty is that the diagram of Fig. \ref{fig:mesmes} for the meson-meson scattering amplitude, which we discuss here for simplicity, quantitatively accounts for the different values of say the $\rho\pi$ and $\phi\pi$ total cross-sections. 
\FIGf
Essentially, as follows from work of Carlitz, Green and Zee \cite{CGZ}, this amounts to noticing that viewed from the $t$-channel the diagram of Fig. \ref{fig:mesmes} corresponds to the following picture. The two mesonic open strings 2 and 3 merge into a unique open string, which then closes up into a closed string, which propagates and then opens up into an open string and ultimately breaks up into the open strings 1 and 4. This picture is presented in Fig. \ref{fig:string}. 
\FIGg
We see that the Pomeron is different because it corresponds to a closed string which couples differently to $\rho$ mesons and to $\phi$ mesons. This is so, because, as we just saw, the Pomeron couples by first opening up into a string that has non-strange quarks at its ends in the case of the $\rho$ and strange quarks at the ends in the case of the $\phi$. The energy denominator associated with these two cases are different and it is clear that the larger denominator will appear in the $\phi$ case since strange quarks are heavier than non-strange quarks and as such the difference between the Pomeron and vector meson trajectories' intercepts is larger in the case of the $\phi$. We immediately understand that the $\phi\pi$ total cross-section will be suppressed with respect to the $\rho\pi$ total cross-section and the suppression factor is calculable and turns out to be $(m_\rho/m_\phi)^2\simeq 0.57$. Now returning to the experimentally more interesting scattering on a proton target, let us concentrate on $J/\psi p$ scattering differential cross-section which can be deduced using vector meson dominance from the measured   $J/\psi$-photoproduction cross-section. The prediction of Carlson and Freund \cite{CF},
$\frac{d\sigma_{J/\psi p}}{dt} / \frac{d\sigma_{\rho p}}{dt}=\frac{m_{\rho}^4}{m_{J/\psi}^4}\simeq 0.0038$, was experimentally confirmed \cite{GLR}.

In the meantime Veneziano's 4-point amplitude was generalized to N-point amplitudes, and people started calculating loop amplitudes. The diagram of Fig. \ref{fig:mesmes} could now be calculated, and this was done by Frye and Susskind \cite{FS} and by Gross, Neveu and Scherk and Schwarz \cite{GNSS} and its careful analysis by Lovelace \cite{L} revealed an amplitude with singularities which violate the well-known analyticity properties of scattering amplitudes. Lovelace noticed that this serious problem could be avoided for bosonic strings, if spacetime had precisely 26 instead of 4 dimensions. Critical string theory was thus started. Schwarz \cite{JHS} and Goddard and Thorn \cite{GT} then found that for superstrings the critical dimension was reduced to 10. 

The advent of QCD put strings on the back burner, until Schwarz and Scherk \cite{SS} made the bold proposal that string theory is much more than a theory of hadrons; it contains the graviton and could be the ultimate physical theory. Then in 1984, things started moving, but that is no longer early string theory history.


 \end{document}